\documentclass[aoas]{imsart}

\usepackage{amsmath,amsfonts}
\usepackage{graphicx,psfrag,epsf}
\usepackage{enumerate}
\usepackage{natbib}
\usepackage{url} 
\usepackage{color}
\usepackage{bm} 

\begin{document}

\begin{frontmatter}

\title{Estimating Scale Discrepancy in Bayesian Model Calibration for ChemCam on the Mars Curiosity Rover}
\runtitle{Model Calibration for ChemCam}

\begin{aug}
\author{\fnms{K. Sham} \snm{Bhat}\thanksref{m1}\ead[label=e1]{bhat9999@lanl.gov}}
\and
\author{\fnms{Kary} \snm{Myers}\thanksref{m1}\ead[label=e2]{kary@lanl.gov}}
\and
\author{\fnms{Earl} \snm{Lawrence}\thanksref{m1}\ead[label=e3]{earl@lanl.gov}}
\and
\author{\fnms{James} \snm{Colgan}\thanksref{m2}\ead[label=e4]{jcolgan@lanl.gov}}
\and
\author{\fnms{Elizabeth} \snm{Judge}\thanksref{m3}\ead[label=e5]{bethjudge@lanl.gov}}

\affiliation[m1]{Statistical Sciences, Los Alamos National Laboratory}
\affiliation[m2]{Theoretical Division, Los Alamos National Laboratory}
\affiliation[m3]{Chemical Diagnostics and Engineering, Los Alamos National Laboratory}

\runauthor{Bhat et al.}
\end{aug}

\begin{abstract}
The Mars rover Curiosity carries an instrument called ChemCam to determine the composition of the soil and rocks. ChemCam uses laser-induced breakdown spectroscopy (LIBS) for this purpose. Los Alamos National Laboratory has developed a simulation capability that can predict spectra from ChemCam, but there are major scale differences between the prediction and observation. This presents a challenge when using Bayesian model calibration to determine the unknown physical parameters that describe the LIBS observations. We present an analysis of LIBS data to support ChemCam based on including a structured discrepancy model in a Bayesian model calibration scheme. This is both a novel application of Bayesian model calibration and a general purpose approach to accounting for such systematic differences between theory and observation in this setting.
\end{abstract}


\end{frontmatter}

\section{Introduction}
\begin{figure}[t!]
\begin{center}
\includegraphics[width=4in]{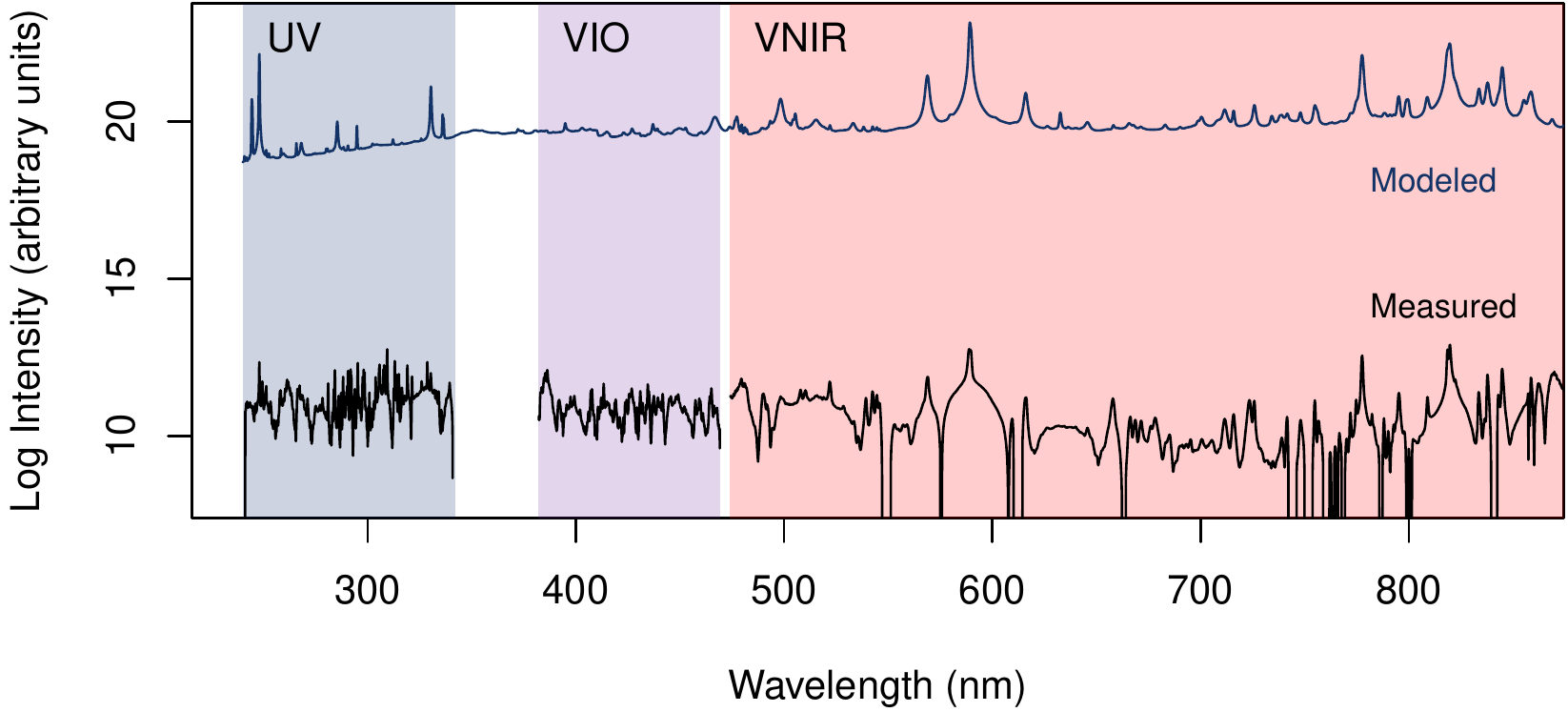} 
\caption{Simulated (modeled) and measured chemical spectra of the compound NaCl over the three ChemCam spectrometers: UV, VIO, and VNIR. Note that we have plotted these spectra on the log scale to aid our analyses, while the spectroscopy community always views spectra on the linear scale.  ChemCam measures chemical spectral intensity in photon counts (bottom line) while ATOMIC computes spectral intensity in power per volume per photon energy per unit solid angle (top line). This results in clear scaling differences, which require quantitative estimation for proper statistical calibration.}
\label{fig:twospect}
\end{center}
\end{figure}

The Mars rover Curiosity was designed to study whether Mars ``ever [had] the right environmental conditions to support small life forms'' \citep{CuriosityWebsite}. As part of the mission, Curiosity carries an instrument called ChemCam, developed by Los Alamos National Laboratory and L'Institut de Recherche en Astrophysique et Plan\'etologie, to determine the composition of the soil and rocks. ChemCam uses laser-induced breakdown spectroscopy (LIBS) for this task. In LIBS, a laser is fired at a target to produce a high-temperature plasma. As the plasma cools, the target emits a spectrum of light over a range of wavelengths that is recorded by a CCD camera. On ChemCam, these are captured with three spectrometers covering three different wavelength ranges: ultraviolet (UV), violet (VIO), and visible and near-infrared (VNIR). Figure \ref{fig:twospect} shows examples of simulated and measured spectra. Each spectrum shows the intensity of light as a function of wavelength. The spectral patterns can be used to identify chemical species and their relative abundances in the target. The presence or absence of certain species and their relative abundances are important clues in answering questions about whether Mars could have ever sustained simple life.

Estimating the chemical composition of soils and rock via LIBS can be difficult. While experts can often easily identify the presence of chemical constituents based on the presence of certain peaks in a spectrum, identifying the relative abundances of the constituents is more difficult due to interactions within the plasma between atoms of the constituents. These interactions, called {\it matrix effects} \citep{Judge16}, can change peak heights in a nonlinear manner. Matrix effects make the disaggregation problem difficult because the spectrum for a target with several chemical species is not a simple linear combination of the spectra for the individual species. Los Alamos National Laboratory has developed a physics simulation code called ATOMIC \citep{magee2004alamos} that can predict spectra in the presence of matrix effects given inputs such as the chemical species, their relative abundances, and others that we discuss later. Our long term goal is to use this simulation capability along with Bayesian model calibration techniques \citep{kennedy:bcc} to determine which chemical species are present in a target and in what relative abundances based on measured spectra, providing vital clues to answering questions about the history of Mars.

This work focuses on the case of estimating plasma temperature and density for simpler compounds. This lets a addres one particular challenge within this larger problem: accounting for scale differences between simulation and observation. LIBS instruments measure chemical spectra in photon counts as a function of wavelength. ATOMIC provides spectra in power per volume per photon energy per unit solid angle. In other words, ATOMIC can predict the shape of the spectrum as a function of inputs. Measured spectra will differ from an ideal simulation by some unknown scaling as seen in Figure \ref{fig:twospect}. That scaling can depend on many factors, such as laser power, standoff distance and angle, and CCD camera properties. Some of these differences can be corrected based on scientific knowledge and testing. In the context of ChemCam, the scientists typically treat this scaling factor as approximately constant within each spectrometer, but it may vary as a function of wavelength. 

Current efforts to account for the scaling factor for LIBS spectra do not take advantage of advanced statistical techniques \citep{Colgan15}. This paper presents an approach for estimating scaling factors within Bayesian model calibration by including them in a structured discrepancy model. The structure allows us to strongly incorporate expert knowledge and avoid problems arising from overly flexible discrepancy specifications \citep{brynjarsdottir2014learning}.

Differences between theory and observation of this type are common. \cite{kennedy:bcc} initially include an ``unknown regression parameter'' that they call $\rho$ that scales the simulation output. This term is typically dropped in later work in this area, apparently for identifiability reasons. One exception is found in \cite{joseph2009statistical}, which uses this term in a similar manner to the one we present. Our scaling difference is a simple instance of a common issue in model calibration: data must be processed because the experiment and simulation do not produce the same output. One example arises in materials science when split Hopkinson pressure bar experiments are used to calibrate flow stress models \citep{sjue2019fast}, but only after the measured strains are converted to stress-strain curves \citep{gray2000classic}. When done outside of the statistical framework, this process can introduce bias or conceal uncertainty. Our work addresses this by including the difference between theory and observation as a structured discrepancy.

Many people have thought about statistical modeling that includes additional components beyond those that account for  the physical process under consideration, and of particular interest here are those that account for effects of the measuring instrument or process. In the case of Raman spectroscopy,  \cite{ray1997simplified} present an approach for determining this instrument effect. The problem also arises in astronomical observations of spectra, as in \cite{van2001analysis}, \cite{lee2011accounting}, and \cite{meng2018conducting}. Statistical quality control methods have been applied to monitor the stability of instrument effects in chromotagraphy \citep{stover1998statistical}. The impact of the detector or sensor on the measured data has also been considered an important source of variability that must be accommodated in statistical modeling under the names of ``machine characteristics and performance" in functional magnetic resonance imaging \citep{genovese1997estimating} and ``systematic variation" in electrophoresis imaging \citep{sellers2007lights}. We note that these effects of the measuring instrument or process are often called {\em instrument response}, but we avoid that term here because ``instrument response'' in the  LIBS community refers to a specific effect that is removed during preprocessing. The LIBS version of instrument response could be included in the framework that we present here, but for now we rely on the scientists' preprocessing. 

In this paper, we present a novel application of Bayesian model calibration to the problem of laser-induced breakdown spectroscopy. This includes an approach to incorporating scale differences between theory and observation into inference based on computationally intensive forward physics models. We use the Bayesian model calibration framework described by \cite{kennedy:bcc} and \cite{higdon2008cmc}. Those papers discuss the concept of a discrepancy model that can be used to capture the systematic differences between the simulation and reality. We will use a highly structured discrepancy to capture the systematic scale differences. We will consider two versions of such a discrepancy --- one constant, one linear with wavelength --- and later discuss a general approach to this problem. Our work will be developed in the context of laboratory measurements that mimic Martian conditions using a ChemCam instrument like the one deployed on the Curiosity rover.

The rest of this paper is organized as follows.   In Section \ref{sec:LIBS}, we give an overview of measured LIBS spectra from the ChemCam instrument and modeled spectra from the ATOMIC computer model.  In Section \ref{sec:calibscfact}, we provide some background on Bayesian model calibration and present our approach for including scaling factor estimation into model calibration. Next, we demonstrate the approach using perfect model experiments in Section \ref{sec:results} and show results on measured ChemCam data from a laboratory experiment. 
Finally, we conclude with some discussion and avenues for future research.

\section{Laser Induced Breakdown Spectroscopy}\label{sec:LIBS}         
  
Here, we give some details on the collection of measured spectra using ChemCam and predictions from the theoretical ATOMIC model. In both the experimental and theoretical case, the output is dependent on the chemical composition of the material being measured or simulated. In the long term, our goal is to automate the identification of the chemical species comprising the target material. For the present, our goal is narrower: build a modeling approach that can correctly identify unknown scaling differences between theory and measurement, while also correctly estimating a number of other physics parameters. As such, we will not be including a detailed discussion of how the chemical constituents affect measurements, nor will we be varying this important input in the theoretical predictions. 

\begin{figure}[t!]
\centering
\includegraphics[width=\textwidth]{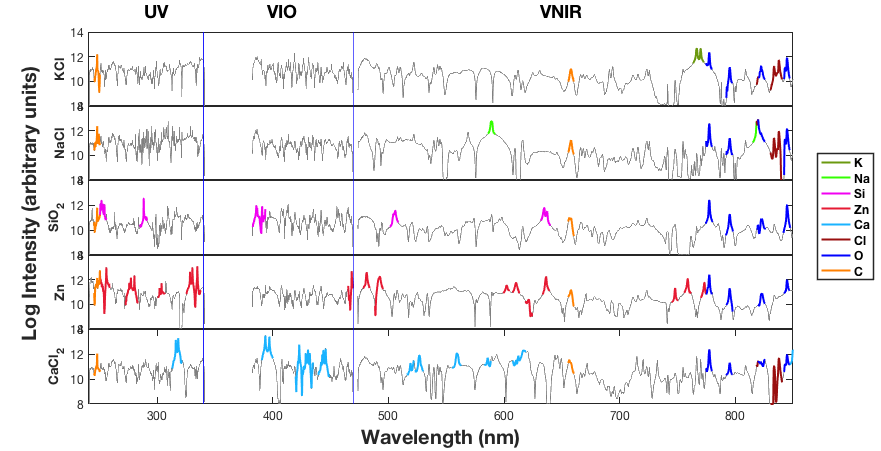}
\caption{Each row shows the observed spectrum for one of our five compounds, computed by averaging the 75 laser shots for that compound. The colored regions indicate the expert-identified elemental peaks that were considered to be the most relevant for these five compounds. These selected peaks are overlaid over the full spectrum in each panel (gray). The blue vertical lines indicate the wavelength break points (at 342 nm and 470 nm) that divide the spectrum across the three instruments: UV, VIO, and VNIR spectrometers. We will use the selected peaks rather than the full set of wavelengths in each spectrum to estimate the scaling factors and physics parameters in each of the three ChemCam spectrometers.} 
\label{fig:elempeaks}
\end{figure}

All of our data, both measured and theoretical, will come from five relatively pure compounds: KCl, NaCl, SiO$_2$, Zn, and CaCl$_2$. For both theory and measurement, rather than working with the full spectra across all wavelengths, we will focus on the wavelengths around an expert-identified set of elemental peaks that are relevant for those five compounds. These peaks are used by experts to identify the presence of these compounds in LIBS spectra, and we believe that the remaining wavelengths outside these peaks will be relatively uninformative for our task. The wavelength locations of these important spectral peaks are known to considerable precision, but observed peak locations can be slightly shifted due to spectral adjustments and the resolution of the detector. For a selected peak at a specified wavelength, we consider a surrounding width that should roughly cover the full width of the peak at half its maximum, but no more than one nanometer (nm) at each side. These peaks and widths are indicated by the colors in Figure \ref{fig:elempeaks}. 

\subsection{Measured Spectra from the ChemCam Instrument}  \label{subsec:chemcamobs}

ChemCam measurements begin by firing a laser at a target, such as soil or rock. The laser ablates a tiny portion of the target and produces a small plasma. This excites atoms and emits photons.  The light from these photons is optically collected by the CCD camera. A diffraction grating refracts the light, separating the wavelengths and directing the light to the appropriate spectrometer: UV (wavelengths 240-342 nm), VIO (382-470 nm), and VNIR (474-850 nm).  The result is a chemical spectrum which is a set of photon counts as a function of wavelength. Different chemical compositions in the target produce different high intensity spectral peaks.  Another source of peaks is the surrounding atmosphere, an unknown proportion of which couples into the plasma, introducing additional peaks into the spectrum. For instance, the Martian atmosphere is 95\% CO$_2$ at a pressure of 7 Torr, and this usually results in a noticeable carbon peak at 247 nm regardless of the composition of the target. More information regarding the details of the collection of the ChemCam experiment may be found in \cite{wiens2013} and \cite{maurice2012}.

Our experimental data, shown in Figure \ref{fig:elempeaks}, come from laboratory measurements collected with a ChemCam instrument under Martian conditions as follows. For each target, laser shots are repeated 30 times at each of three locations on the target. We discard the first five shots at each location as they are potentially contaminated by surface dust, and are left with 75 measured spectra for each target. These are post-processed by the scientists to remove the so-called dark spectra (what the spectrometer records with no light hitting it) and other effects. We examined the shot-to-shot variation in the data and judged it to be small and lacking any structure. Therefore our observation for each target compound (KCl, NaCl, SiO$_2$, Zn, and CaCl$_2$) is the average of the 75 post-processed spectra for that compound. 

\subsection{ATOMIC Computer Model}  \label{subsec:atomic}

The ATOMIC forward model was developed to simulate the emission spectra of chemical compounds using first principles theoretical atomic physics, the details of which can be found in \cite{magee2004alamos}. 
Briefly, ATOMIC is a general purpose plasma modeling and kinetics code that has been designed to compute emission (or absorption) spectra from plasmas either in local-thermodynamic equilibrium (LTE) or in non-LTE. The primary inputs are the plasma temperature and density, and a model describing the atomic structure and scattering data of the constituent material(s).
ATOMIC receives data (energy levels, transition probabilities, quantum numbers, etc.) from the Los Alamos suite of atomic physics codes \citep{Fontes15}. In the simulations discussed in this paper, the results were generated from the \textsc{CATS} code \citep{cowan} with modifications made for plasmas generated from LIBS \citep{Colgan14}. ATOMIC then uses these data to compute the average ionization of the plasma (for a given temperature and density) and the resulting emissivity of the plasma.  ATOMIC has recently been used to model the emissivity from a number of LIBS plasmas \citep{Colgan15,Colgan14,Judge16}.

For each of our five compounds, we run simulations over the space of three parameters:
\begin{itemize}
\item plasma temperature $T$, with units electron volts (eV) and range $[0.5,1.5]$
\item mass density $\rho$, with units of g/cm$^3$, and range $[-7, -4]$ on the $\mbox{log}_{10}$ scale
\item proportion of target $p$ as opposed to atmosphere in the plasma, with range $[0.1,0.98]$
\end{itemize}
This gives a total of 15 parameters (three parameters for each of five compounds). These ranges also indicate the bounds for our uniform prior distributions in the calibration described below to estimate these parameters. The output is a spectrum for the specified compound with intensity as a function of wavelength. The simulation provides intensity as power per volume per photon energy per unit solid angle. ATOMIC predictions account for matrix effects but do not account for any effects arising from the spectrometer itself. Computation time for a single run of ATOMIC for a compound can range from minutes to hours on a high-performance computing system, depending on the chemical complexity of the compound.

\section{Calibration with Scale Differences}\label{sec:calibscfact}

\subsection{Computer Model Calibration}  \label{subsec:compmodcalib}

Bayesian computer model calibration is, by now, a well studied problem. We will generally follow the approach described in \cite{kennedy:bcc} and extended in \cite{higdon2008cmc}. We will review this briefly. In short, the goal is to find the parameters and systematic biases of a computer simulation that make it best match experimental data.

Assume that an output $y$, possibly a vector, is observed with measurement error from a true physical system (e.g., a ChemCam measurement). Physical reality can be be approximated by a simulator (e.g., ATOMIC) denoted $\eta(\cdot)$ with parameters $\theta$. This approximation may have systematic biases or discrepancy, denoted $\delta$, which may be a function of the measurement index (e.g., a function of wavelength) or other experimental conditions. Therefore, a general model for the measurement $y^{obs}$ using the computer simulation $\eta(\cdot)$ is
\begin{equation}
y^{obs} = \eta(\theta) + \delta  + \epsilon,
\label{eq:data_model_0}
\end{equation}
where $\epsilon$ captures the measurement error. In our case, the observation and simulations consist of the spectra at the selected wavelengths shown in Figure \ref{fig:elempeaks}. The spectra are concatenated across compounds. Thus, the observation vector includes all five compounds, as does each run of the simulation.

The goal in the Bayesian paradigm is to estimate the posterior distribution for $\theta$ and $\delta$. Since ATOMIC, like other simulators used in such problems, is computationally expensive, we need to build an emulator, a statistical approximation to the physics simulation. For this work, we follow the approach of \citet{higdon2008cmc} as implemented in the software package GPMSA \citep{gattiker2016}. The simulator is run over a design of inputs (described later for our specific case). For each input vector $t_i$, we observe the simulation result $\eta_i = \eta(t_i)$. These will be used to build the emulator. The emulator is decomposed as
\begin{equation}
\eta(t)= \mu + \sigma \sum_{h=1}^{p_\eta} k_h w_h(t),
\end{equation}
where $\mu$ is the mean vector of the training simulations and $\sigma$ is a scalar computed as the standard deviation across the output index over all the training runs. The $k_h$ are computed based on a singular value decomposition of the standardized simulations. Let $z_i$ be the $i^{th}$ training simulation standardized with $\mu$ and $\sigma$. Let $Z = [z_1, \cdots, z_m]$ be the matrix with the standardized training simulations as columns. Compute the singular value decomposition: $Z = USV'$. Let $K = US / \sqrt{m}$. The $k_h$ are the columns of this matrix $K$, which may be truncated to include only those associated with large singular values. We typically choose $p_{\eta}$ to account for at least 99\% of the total variance in the training simulations. We project $Z$ onto $K$ to get the training weights for each basis vector. The weights $w_h(t)$ for each basis vector are fit using a Gaussian process (GP) over the input design. Hyperpriors for the GP, as well as further details on emulation, are described in \cite{higdon2008cmc}.

The discrepancy model $\delta$ is also based on a basis representation, 
\begin{equation}
\delta= \sum_{h=1}^{p_\delta} d_h \alpha_h.
\end{equation}
In \cite{higdon2008cmc}, the basis vectors $d_h$ are normal kernels over the output space (wavelength in our case), and the basis weights $\alpha_h$ have a zero mean Gaussian prior with a marginal precision $\lambda_{\alpha h}$. If the discrepancy is modeled as a function of experimental conditions, the prior becomes a Gaussian process over this space.  The discrepancy can be viewed as a convolution of the kernels $d_h$ and the weights $\alpha_h$. A weakly informative Gamma(1, 0.001) prior is placed on the $\lambda_{\alpha h}$. Unless the data informs otherwise, $\lambda_{\alpha h}$ will remain at a large value consistent with almost zero discrepancy.   Lastly, the measurement error is represented as $\epsilon_i \sim N(0,\frac{1}{\lambda_{yi}} \Sigma_{yi}$), where $\Sigma_{yi}$ is an $n \times n$ covariance matrix that may be specified by the user. 


\subsection{Discrepancy for Scale Differences} \label{subsec:implem}

As seen in LIBS and other applications, the observations are a scaled version of the output from the physics simulations at some unknown value of the input. In other words, we expect the scaling factor to be multiplicative. Further, after some exploratory data analysis, we decided to build the emulator on the log of the output. The output can vary by orders of magnitude of the design space, and the emulator worked empirically better on the log scale. For this reason, we let $y^{obs}$ be the log of a measured spectrum. On the log scale, the multiplicative scaling factor becomes an additive discrepancy. 

While the framework of \cite{higdon2008cmc} provides considerable flexibility, we prefer to constrain any possible discrepancy for both statistical and scientific reasons. Statistically, a flexible discrepancy can cause identifiability problems. A spectrum is composed of a number of peaks. If a discrepancy model over wavelength can make adjustments at the scale of peak widths, then this will be confounded with physics model adjustments to peak heights. For this reason, we need to carefully chose a discrepancy basis that adjusts physics model output on scales that avoid confounding. We also need to ensure that our discrepancy basis is not confounded with parameter effects in the physics code; this is discussed more below.

Scientifically, we expect discrepancy to behave in particular ways, so we should encode that knowledge in a constrained basis. First, as mentioned earlier, the scaling factor is mostly assumed to be constant within each of the spectrometers. However, there are a few other possibilities of scientific interest. In particular, there are well-known LIBS instrument effects at the edges of each of the spectrometers where the sensitivity decreases. These are known well enough to be corrected in post-processing \citep{wiens2012chemcam}, but effects may linger. In general, the scaling factor could change as a function of wavelength. For this paper, we will consider two simple models: a constant scaling factor for each spectrometer and a scaling factor for each spectrometer that varies linearly with wavelength. The latter is the simplest model that considers variation with wavelength. In Section \ref{sec:discussion}, we discuss extensions of this approach when the modeling is done on the linear scale.

For the constant scaling factor model, we use three basis functions for the discrepancy, one for each spectrometer,
\begin{equation}
\delta_{con} =  d_{UV} \alpha_{UV} + d_{VIO} \alpha_{VIO} + d_{VNIR} \alpha_{VNIR}.
\end{equation}
Each of the basis vectors $d_{UV}$, $d_{VIO}$, and $d_{VNIR}$ has length equal to the number of wavelengths in our model. We set the $\omega$-th entry $d_{h, \omega}=1$ for all wavelengths $\omega$ in spectrometer $h$ and $0$ otherwise.  Hence the basis weights $\alpha_k$ may be interpreted directly as scaling factors.  

For the linear scaling factor model, we use six basis vectors for the discrepancy representing the intercept and slope for each of the three spectrometers.
\begin{eqnarray}
\delta_{lin} & = & d_{UV,0} \alpha_{UV,0} + d_{UV,1} \alpha_{UV,1} \\
                  & + & d_{VIO,0} \alpha_{VIO,0} + d_{VIO,1} \alpha_{VIO,1} \nonumber \\
                  & + & d_{VNIR,0} \alpha_{VNIR,0} + d_{VNIR,1} \alpha_{VNIR,1}. \nonumber 
\end{eqnarray}
Again, all of the basis vectors have length equal to the number of wavelengths in our model. The intercept terms $d_{h,0}$ are identical to the $d_{h}$ from the constant model. For the slope terms, we set the $\omega$-th entry $d_{h,1,\omega} = \omega$ when the wavelength $\omega$ is in spectrometer $h$ and $0$ otherwise. Now, the basis weights describe a linear model for the scaling factor.

Returning to the question of identifiability, we believe that it is especially important to evaluate the parameter effects using the emulator. In our experience in other computer model calibration problems, there are often parameters whose effect is largely in the form of a constant shift or a change in the slope of an output. Therefore, we run the risk that our discrepancy model may mimic the main effect of a physics model parameter. To explore this possibility, we computed a simplified main effect for each parameter in which each physics parameter is varied over its range while all other parameters are fixed at the center of the design range. This is good practice under any circumstance, but especially so here. We present the results of this diagnostic study in Section \ref{subsec:emdiag}.

Prior distributions for the ATOMIC model parameters are assumed to be uniform on the parameter bounds given in Section \ref{subsec:atomic}. We use $p_\eta = 15$ principal component basis vectors to build the emulator. We construct the measurement error matrix $\Sigma_{yi}$ by treating wavelengths independently --- based on the assumption of independent Poisson error in the photon counts for each wavelength --- and taking the variance over the 75 shots at each wavelength.  We estimate the error precision terms $\lambda_{yi}$.

\section{Results} \label{sec:results}

\begin{figure}[t!]
\begin{center}
\includegraphics[width=3.25in]{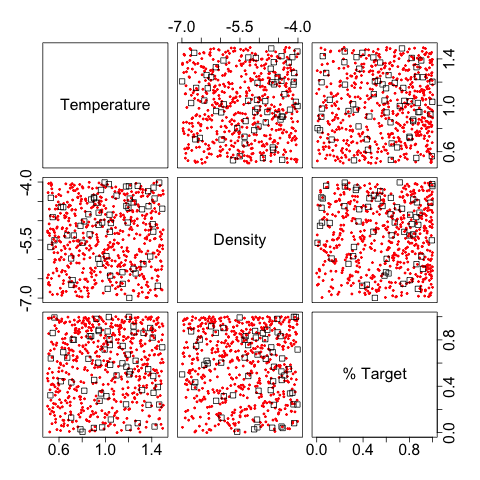} 
\caption{Bivariate scatterplot for an $m = 600$ point LHS design in red, showing a subset of the full design.  Here we show the three ATOMIC model parameters $T$, $\rho$, and $p$ for one compound, KCl. This is a subset of the full 600-run by 15-parameter design covering the parameters for all five compounds.  The black squares show a separate set of thirty input settings chosen for emulator validation. }
\label{fig:lhsdesign1}
\end{center}
\end{figure}

Here, we present results and diagnostics from the calibration of ATOMIC spectra to ChemCam spectra to learn about the scaling factors for each of the three spectrometers.  For this entire study, we use two sets of ATOMIC runs, a subset of which is shown in Figure \ref{fig:lhsdesign1}, for emulator construction and evaluation. The red points are a 600-run Latin hypercube design for training. The black squares are a 30-run Latin hybercube design for testing and evaluation. This figure shows the three parameters for KCl, which are a subset of the full 600-run, 15-parameter design covering the parameters for all five compounds. For all estimation, we use MATLAB software called GPMSA \citep{gattiker2016} for emulation and model calibration. GPMSA employs Markov chain Monte Carlo (MCMC) methods to estimate unknown parameters.  For the calibration for each of the three spectrometers, we ran the MCMC for 40,000 iterations, after a burn-in of 3,750 samples using an adaptive step.  

Below we first present the results for the emulator diagnostics using cross-validation and main effects.  Next we show the calibration results for a perfect model experiment for both the constant and linear scaling factors, and last we give the results for our calibration of the ATOMIC model to the ChemCam LIBS data.

\subsection{Emulator Diagnostics}  \label{subsec:emdiag}

We begin by evaluating the quality of the emulator using the test set. To do so, we compute $R^2 = 1 - \sigma^2_{res} / \sigma^2_{raw}$ where $\sigma^2_{raw}$ is the variance of the test set around its empirical mean and $\sigma^2_{res}$ is the variance of the test set residuals, i.e., the test set minus the emulator predictions. Computed pointwise at each of the wavelengths that we modeled, the minimum $R^2$ was 0.9904, suggesting that the emulator does an excellent job.

\begin{figure}[t!]
\begin{center}
\includegraphics[width=\textwidth]{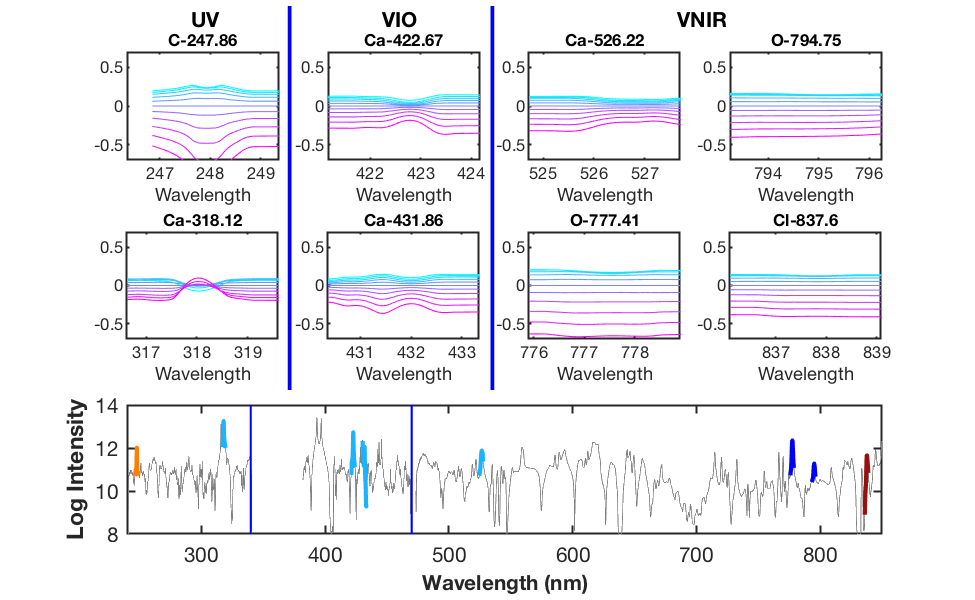}
\caption{Of our 15 parameters (3 input parameters $\times$ 5 compounds), target proportion $p$ for CaCl$_2$ appears to be the most likely to be confounded with our discrepancy, and we explore that in this figure. The bottom panel shows the measured spectrum for CaCl$_2$ with the 8 selected peaks colored as in Figure \ref{fig:elempeaks}. The 8 panels in the top two rows show the main effect plots for the target proportion of CaCl$_2$ for these peaks, presented in increasing order of wavelength and labeled with the element and center wavelength.  For each panel, we fix the parameters plasma temperature $T$ and density $\rho$ at the center of their ranges and vary the target proportion parameter $p$ over its domain, with blue indicating the lower bound and pink the upper bound. We subtracted the center prediction from all results. Even here, we expect there to be no problem with identifiability because the carbon peak at 247.86 nm has additional structure, and the magnitudes of the other shifts are neither constant across all peaks nor varying with any notable structure, linear or otherwise.}
\label{fig:maineff1}
\end{center}
\end{figure}

As discussed earlier, the main effects of the parameters are especially important in this application because of potential identifiability issues with the discrepancy. In particular, we will be concerned if we find parameter effects that cause nearly constant shifts or shifts that are approximately constant and change linearly over length. Our explorations found that the most worrisome parameters in our study are those associated with target proportion $p$, something we will discuss later in Figure \ref{fig:pfmpost}. Across our five compounds, the effect of $p$ for CaCl$_2$ showed the most potential difficulty as illustrated in Figure \ref{fig:maineff1}. Even here, there seems to be little cause for concern. Although many of the peaks, such as the three oxygen peaks, show roughly constant shifts, other peaks, particularly the carbon peak in the top-left panel, can be used to separate these effects from discrepancy. Even among peaks that show constant shifts, the magnitudes of those shifts are neither constant nor varying with discernible structure across wavelengths.

\subsection{Perfect Model Experiment}  \label{subsec:pfm}

To demonstrate that computer model calibration approaches can be successful in estimating the ATOMIC model parameters and the scaling factors for chemical spectra output, we perform several perfect model experiments.  From the 600 runs that were used to train the emulator, we hold out one run $\theta^\ddag$ and add white noise and the relevant scaling factor (constant or linear) to its modeled spectrum $\eta(\theta^\ddag)$ to create a synthetic measured spectrum $y^\ddag$.  We apply our calibration approach to $y^\ddag$ to verify that we can recover its original parameters $\theta^\ddag$ and true generating discrepancy.  We repeated this process for 12 different runs selected from interior design points that are closest to the center of the design,  while not having a value in the top or bottom 5\% of the range for any parameter. 

\begin{figure}[t]
\begin{center}
\includegraphics[width=\textwidth]{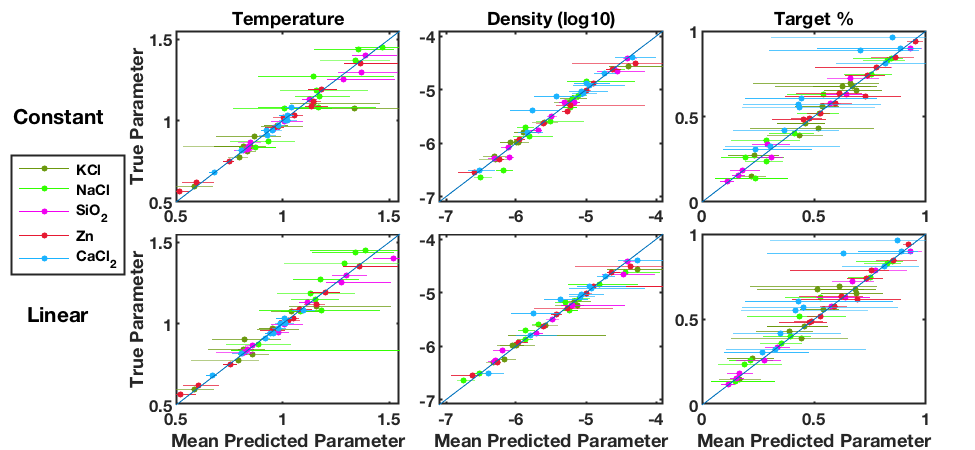}
\caption{Plots of the posterior mean vs. the true value of the ATOMIC model parameters for 12 perfect model experiments. Each panel has 60 points: 12 experiments times five compounds.  The top row represents the constant scaling factor case, the bottom row the linear scaling factor.  The columns from left to right represent the temperature, density, and target proportion parameters for all five compounds.    The 95\% credible regions shown with horizontal lines  largely cover the true value of the parameter (when the credible interval crosses the diagonal identity line). 
 }
\label{fig:pfmpost}
\end{center}
\end{figure}

\begin{figure}[t!]
\begin{center}
\includegraphics[width=\textwidth]{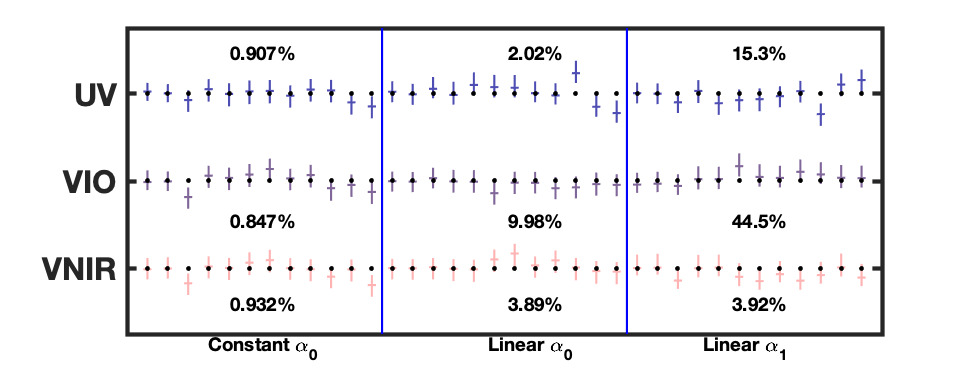}
\caption{For each spectrometer and scaling factor parameter, we show the posterior mean and 95\% credible interval for all 12 perfect model experiments. These intervals are calculated by 1) computing the mean and .025 and .975 quantiles from the MCMC sample, 2) recentering the 95\% interval by subtracting the true parameter value, and 3) standardizing the width of the interval by the average of the 12 standard deviations from each MCMC sample. For most experiments, the true scaling parameters are accurately recovered, as indicated by the intervals covering the black dots. To quantify the size of our uncertainty, we report the average length of the 95\% interval over the 12 experiments as a percentage of the true parameter value.  The slope parameter for the VIO spectrometer has the largest uncertainty; the average length of its 95\% interval is 44.5\% of the true value. This uncertainty could be due to the relative paucity of data for this spectrometer, as shown in Figure \ref{fig:elempeaks}. }
\label{fig:pfmpostIR}
\end{center}
\end{figure}

The results, shown in Figures \ref{fig:pfmpost} and \ref{fig:pfmpostIR}, are encouraging. Figure \ref{fig:pfmpost} shows that we are accurately able to recover the true ATOMIC parameters, particularly once posterior uncertainty is taken into account. The estimates for temperature and density are particularly close to the true values, while we have wider uncertainties for our estimates of target proportion. Figure \ref{fig:pfmpostIR} shows that we also accurately recover the discrepancy parameters used to generate the data for each of the perfect model experiments. We note that the slope parameters of the VIO spectrometer exhibit wide uncertainty (noted 44.5\% in the figure). This is likely related to the paucity of data for this spectrometer --- there are just 16 peaks in the VIO spectrometer as seen in Figure \ref{fig:elempeaks}.

\subsection{ChemCam LIBS Data}  \label{subsec:chemcam}
\begin{figure}
\begin{center}
\includegraphics[width=\textwidth]{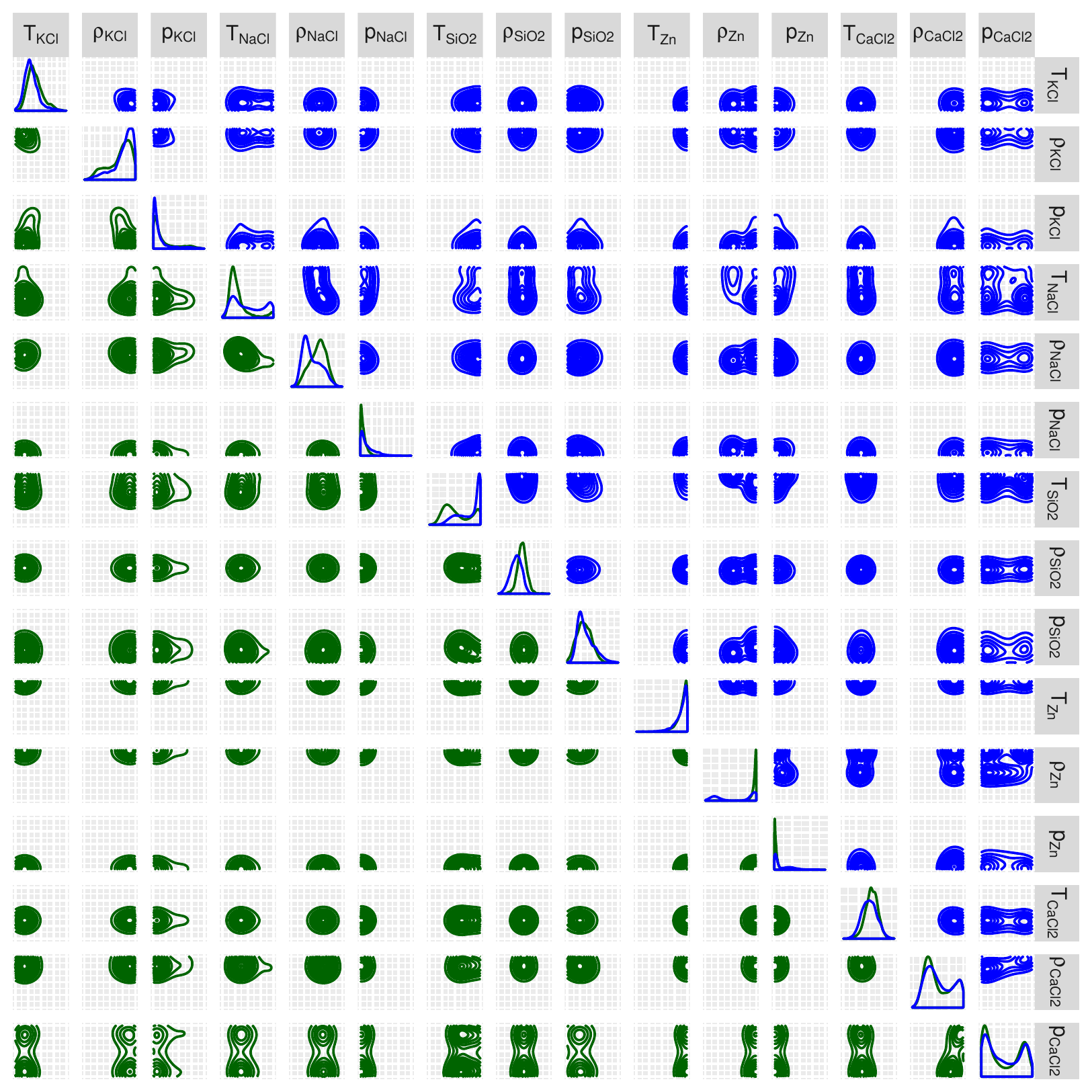}
\caption{Posterior distributions for the ATOMIC model parameters for the constant (green) and linear (blue) scaling factors.  The parameters are temperature (T), density ($\rho$), and target proportion (p) for each of the five compounds: KCl,  NaCl, SiO$_2$, Zn, and Ca$_2$Cl.}
\label{fig:fullpost1}
\end{center}
\end{figure}

These perfect model results and our earlier diagnostics strongly suggest that we can accurately model measured ChemCam data, which we now consider. Figure \ref{fig:fullpost1} summarizes the posterior distributions for the ATOMIC model parameters for both the constant (green) and linear (blue) scaling factors. There are some visible differences between the two models.
For instance the model with the constant scaling factor favors a smaller value for the NaCl temperature parameter, while the linear scaling model is bimodal and generally flatter. There is a slight tendency for this behavior in other parameters too (e.g., the Zn parameters). This might indicate that the more flexible linear scaling model is inducing fewer constraints on the physics model, but this effect is not pronounced or always consistent (e.g., SiO$_2$ temperature). A few parameters (e.g., the CaCl$_2$ density and target proportion) show evidence of bimodality, indicating that ATOMIC may match experimental data with more than one parameter setting.

\begin{figure}
\begin{center}
\includegraphics[width=\textwidth]{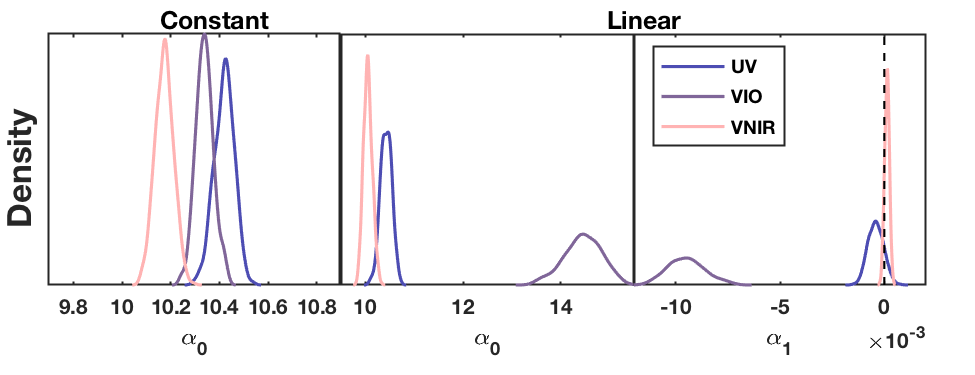}
\caption{Posterior distributions for the  constant scaling parameter (left) and for the intercept term (center) and slope term (right) in the linear scaling model. The UV and VNIR constant scaling parameters are similar in value to the intercepts for the linear scaling model, and the slope parameters span the zero line. The VIO spectrometer shows a strong linear effect. The intercept term for VIO is higher than the constant scaling parameter, and the posterior for the slope is significantly negative.}
\label{fig:irpost}
\end{center}
\end{figure}

The difference between the two scaling factor models is more pronounced for the discrepancy parameters shown Figure \ref{fig:irpost}. For the UV and VNIR spectrometers, the constant and linear scaling factor models are similar. 
The posterior distributions for the intercept term in the linear model (center panel) are close to the posterior distributions for the respective constant terms (left panel). The slope parameters for those two spectrometers (right panel) both span the zero line. The VIO spectrometer, however, seems to strongly favor a linear scaling factor. The intercept term (center panel) is noticeably higher than the constant term (left panel). Further, the slope parameter appears to be significantly negative, with no mass near zero. 
This is a surprise given the scientists' expectation of a constant shift.

%
%

\begin{figure}[t!]
\begin{center}
\includegraphics[width=\textwidth]{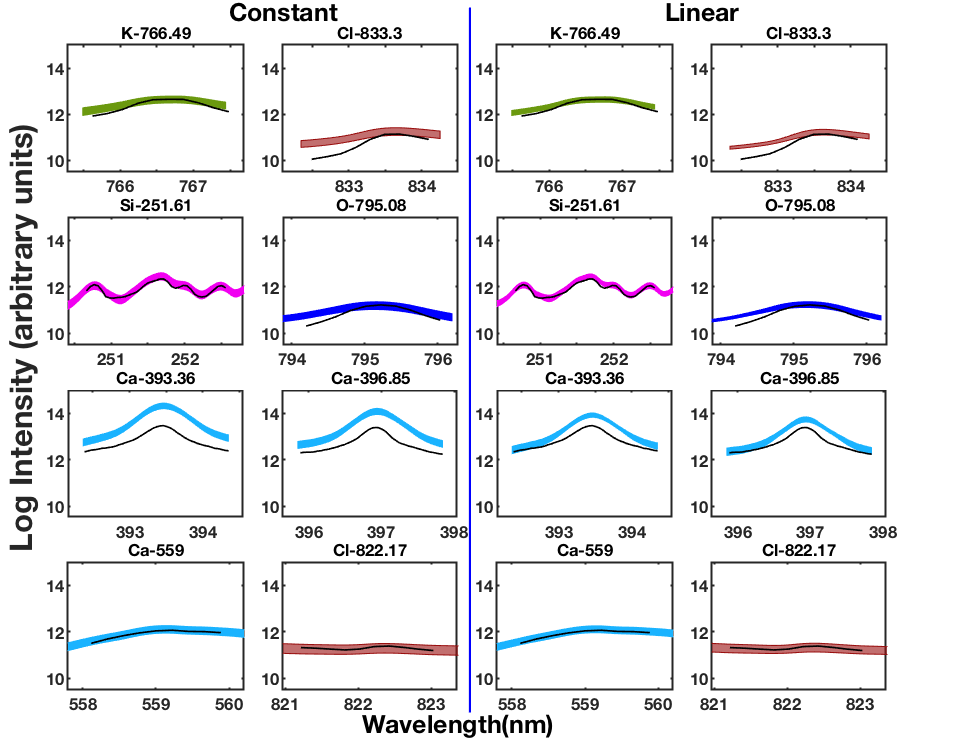}
\caption{Plots of the posterior calibration model predictions and 95\% credible intervals (shaded areas color coded by element similar to Figure \ref{fig:elempeaks}) for the constant scaling factor approach (left) and linear scaling factor approach (right).  Black lines in each panel show the measured spectral peaks. While we analyzed all the selected peaks for each compound, in this figure we highlight a few from KCl (top row), SiO$_2$ (middle row), and CaCl$_2$ (bottom two rows). The panel titles indicate the element and center wavelength of the peak. }
\label{fig:IRpredssing1}
\end{center}
\end{figure}

Figure \ref{fig:IRpredssing1} shows calibrated predictions from the emulator (i.e., use the draws of the parameters from the posterior distribution as inputs to the emulator and add draws from the posterior of the scaling discrepancy) for both scaling models for a subset of peaks from KCl (top row), SiO$_2$ (middle row), and CaCl$_2$ (bottom two rows). The color-coded bands show the 95\% credible intervals for the posterior predictions of the data used in calibration. The black lines show the experimental data. We are looking to see the colored bands mostly cover the black lines. The left panels show the predictions using the constant scaling factor and the right panels show the predictions using the linear scaling factor. 

Overall, the calibrated model does well in capturing the data. In general, the calibrated predictions match the overall level of the response. The calibrated predictions sometimes miss at the ends of the peaks (e.g., O-795.08). This might indicate that ATOMIC has trouble replicating the shoulders of the peaks. This suggests that further discrepancy modeling may be helpful to correct this behavior. 

\begin{figure}
\begin{center}
\includegraphics{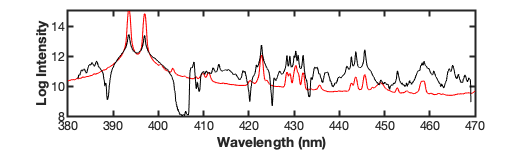}
\caption{Comparison of experimental data (black) and a simulation (red) for CaCl$_2$ on the VIO spectrometer. Here we shifted the simulation using the mean intercept, but not the linear term, for the VIO linear scaling model. The ATOMIC simulation shows a slow decay from the prominent calcium peaks between 390 and 400 nm. This decay is not evident in the experimental data. This difference may be the major contributor to the significant linear effect we found for this spectrometer. Since calcium dominates this spectrometer, the linear scaling effect may be the result of ATOMIC's modeling of CaCl$_2$ and not intrinsic to the spectrometer itself.}
\label{fig:CaCl2VIO}
\end{center}
\end{figure}

The predictions for most peaks are similar between the two models, but the calcium peaks in the third row demonstrate the significance of the linear scaling factor. Both models have difficulty in capturing these peaks, but the linear scaling model does a much better job. Calcium has 12 of the 16 selected peaks in the VIO spectrometer, which is the only instrument to show a significant linear response. Because calcium peaks dominate this spectrometer, this effect may have more to do with this compound than the instrument. As shown in Figure \ref{fig:CaCl2VIO}, the simulated spectra for CaCl$_2$ have broad peaks that decay slowly over the range of this spectrometer. The observed spectrum does not show this behavior, and the linear scaling factor helps to correct the difference. There seems to be a real effect here worth investigating further, whether it's due to the spectrometer's characteristics or to ATOMIC's ability to simulate this compound. 

\section{Discussion}\label{sec:discussion}

Building an effective discrepancy model is a difficult process because of the potential for confounding with the physics model \citep{brynjarsdottir2014learning}. Careful construction of the modeling class and prior distribution is vital to getting valid solutions for calibration. This is especially true in the case when the scaling factor discrepancy doesn't necessarily represent missing physics. In our case, we have no reason to think that ATOMIC is not correctly modeling the underlying physics of LIBS. What we are missing is a forward model of the detector. Treating this as a strongly constrained discrepancy lets us solve the calibration problem without the need for this detector model.

As discussed, we have entertained two fairly simple scaling factor models. One could consider a more general function of wavelength, $f(\omega)$. For instance, this could be crafted to account for edge effects within each spectrometer. Other physics needs can also be represented here. As before, the main effects of the physics parameters should be examined to avoid confounding. The analysis can also be considered on the original scale as opposed to the log scale that we use here. In this case, the model becomes
$
y^{obs} = e^{f(\omega)} \odot \eta(\theta) + \epsilon,~
$
where $\odot$ indicates pointwise multiplication. As discussed earlier, this is similar to the initial model development described in \cite{kennedy:bcc}. Our case is somewhat easier since our multivariate output makes the problem more identifiable. We could also consider a hierarchical approach to either connect discrepancies across spectrometers or to allow scaling to vary somewhat across compounds.

In the future, we will consider compounds made of several elements instead of the simple 2-element compounds considered here. This work will require the estimation of a new set of parameters: the fraction of each element in the target. This aspect of estimation is a key component of addressing the science mission of ChemCam. It remains to be seen how this new parameter space will interact with potential scaling issues. The current work lays the groundwork for this larger goal.

\section*{Acknowledgments}

Research presented in this article was supported by the Laboratory Directed Research and Development program of Los Alamos National Laboratory under project number 20180097ER.

\bibliographystyle{apalike} 
\bibliography{mis}%

\end{document}